\documentclass{revtex4}
\usepackage{epsfig}
\usepackage{amssymb}
\usepackage{amsmath}
\usepackage{amsfonts}
\usepackage{graphicx}
\usepackage{mathrsfs}
\usepackage[dvips]{color}
\usepackage{multirow}

\newcommand{\R}{\mathbb{R}}

\newcommand{\fn}{{\,\mathfrak{n}\,}}

\newcommand{\fz}{\mathfrak{z}}

\newcommand{\cO}{\mathcal{O}}

\newcommand{\be}{\begin{equation}}
\newcommand{\ee}{\end{equation}}
\newcommand{\bea}{\begin{eqnarray}}
\newcommand{\eea}{\end{eqnarray}}
\newcommand{\nn}{\nonumber}
\newcommand{\kt}{\rangle}
\newcommand{\br}{\langle}

\newcommand{\ed}{\end{document}}

\newcommand{\rz}{{\mbox{\scriptsize${\rm Z}$}}}

\newcommand{\bi}{\begin{itemize}}
\newcommand{\ei}{\end{itemize}}

\newcommand{\bce}{\begin{center}}
\newcommand{\ece}{\end{center}}

\newcommand{\sB}{\mathscr{B}}

\newcommand{\sE}{\mathscr{E}}

\newcommand{\RE}{{\rm Re}}
\newcommand{\IM}{{\rm Im}}

\begin{document}

\title{Spectral Singularities and Whispering Gallery Modes of a Cylindrical Gain Medium}

\author{Ali~Mostafazadeh and Mustafa~Sar{\i}saman}
\address{Department of Mathematics, Ko\c{c}
University,\\ Sar{\i}yer 34450, Istanbul, Turkey\\
amostafazadeh@ku.edu.tr}

\begin{abstract}
Complex scattering potentials can admit scattering states that behave exactly like a zero-width resonance. Their energy is what mathematicians call a spectral singularity. This phenomenon admits optical realizations in the form of lasing at the threshold gain, and its time-reversal is responsible for antilasing. We study spectral singularities and whispering gallery modes (WGMs) of a cylindrical gain medium. In particular, we introduce a new class of WGMs that support a spectral singularity and, as a result, have a divergent quality factor. These singular gallery modes (SGMs) are excited only if the system has a positive gain coefficient, but typically the required gain is extremely small. More importantly given any amount of gain, there are SGMs requiring smaller gain than this amount. This means that, in principle, the system lacks a lasing threshold. Furthermore, the abundance of these modes allows for configurations where a particular value of the gain coefficient yields an effective excitation of two distant SGMs. This induces lasing at two different wavelengths.
\medskip

\noindent {Pacs numbers: 03.65.-w, 03.65.Nk, 42.25.Bs, 42.60.Da, 24.30.Gd}
\end{abstract}

\maketitle
\section{Introduction}

Complex scattering potentials have numerous applications in modeling various physical phenomenon \cite{muga}. Among their remarkable properties is that unlike a real scattering potential they can support a peculiar type of scattering states that share the characteristic properties of zero-width resonances \cite{prl-2009}. The underlying mathematical structure responsible for these states is the concept of a spectral singularity \cite{ss}. The physical meaning of spectral singularities has been given recently in \cite{prl-2009}. This has motivated a detailed study of physical aspects of spectral singularities \cite{pra-2009} -- \cite{plyushchay}. Among the remarkable results obtained in this direction is the discovery that in optics a spectral singularity corresponds to lasing at the threshold gain \cite{pra-2011a} while its time-reversal provides the mathematical basis of coherent perfect absorption of light that is also called antilasing \cite{longhi-phys-10,antilasing}.

In Refs.~\cite{pla-2011,prsa-2012}, we explore the optical spectral singularities (OSSs) in the radial (transverse) modes of an optically active media with a spherical geometry. The condition that such an OSS be present puts a lower bound on the size of the active medium. For a homogeneous spherical gain medium made out of a typical dye laser material (Eq.~(\ref{specific}) below), the lower bound on the radius of the sample turns out to be about 3~{\rm mm}, \cite{pla-2011}. Given the difficulties associated with maintaining a uniform gain coefficient throughout such a large sample, we have considered in \cite{prsa-2012} the possibility of reducing the size of the gain medium by using a high-refractive index coating. This turns out to decrease the lower bound on the radius of the sample only by a factor of 3. The impossibility of achieving this kind of OSS in micron size gain media is connected with the fact that they appear in the radial modes. For these modes the Poynting vector of the wave propagating inside the gain medium is along the radial direction and the length of the optical path along which the wave is amplified is essentially determined by the radius of the sample. This observation suggests that it should be possible to create OSSs using much smaller samples, if we consider surface waves propagating along the boundary of a spherical sample. The best-known examples of these are the whispering gallery modes (WGMs), \cite{rayleigh-1910} --\cite{WGM}. Motivated by this idea, we give in this paper a detailed analysis of the problem of spectral singularities for the surface modes of a homogeneous cylindrical gain medium. In particular we explore the possibility of generating spectral singularities in the WGMs of this model. We study a sample with a cylindrical geometry, mainly because the WGMs for a cylindrical medium \cite{rowland,WGM-cyl-acoustic} are easier to deal with than those of a spherical medium.

Consider an optically active material confined in an infinite cylinder of radius $a$.
     \begin{figure}
      \begin{center}
      \includegraphics[scale=.5]{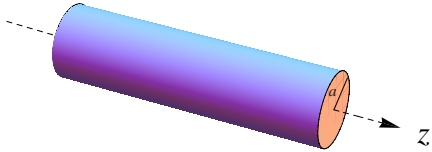}
      \caption{(Color online) An effectively infinite cylindrical gain medium of radius $a$.}
      \label{figure1}
      \end{center}
      \end{figure}
Let $\fn$ denote the complex refractive index of this material with real and imaginary parts, $\eta$ and $\kappa$, so that $\fn=\eta+i\kappa$, and suppose that $\fn$ has a constant value inside the confining cylinder and in time. Then the time-harmonic electromagnetic waves, $\vec \sE(\vec r,t)=e^{-i\omega t}\vec E(\vec r)$ and $\vec \sB(\vec r,t)=e^{-i\omega t}\vec B(\vec r)$, that interact with this system satisfy the Maxwell equations:
    \begin{align}
    &\vec{\nabla}\cdot\vec{ D} = 0, &&
    \vec{\nabla}\cdot\vec{ B} = 0, \label{equation1}\\
    &\vec{\nabla} \times \vec{ E} -i \omega \vec{ B} = 0, &&
    \ \vec{\nabla} \times \vec{ H} + i \omega \vec{ D} = 0,
    \label{equation2}
    \end{align}
where $\omega$ is the angular frequency of the wave,
    \begin{align}
    &\vec{D} := \varepsilon_0\fz(\rho) \vec{E}, &&
    \vec{H} :=\mu_0^{-1} \vec{B},
    \label{D-H}
    \end{align}
$\varepsilon_0$ and $\mu_0$ are respectively the permeability and permittivity of the vacuum, $(\rho, \varphi, z)$ are cylindrical coordinates, and
    \be
    \fz(\rho):=\left\{\begin{array}{ccc}
    \fn^2 & {\rm for} & \rho<a,\\
    1 & {\rm for} & \rho\geq a.
    \end{array}\right.
    \label{e1}
    \ee
Inserting (\ref{D-H}) in (\ref{equation2}) and eliminating $\vec B$ in these equations, we find the Helmholtz equation,
    \begin{equation}
    \left[\nabla^{2} +k^2\fz(\rho)\right] \vec{E}(\vec{r}) = 0,
    \label{equation4}
    \end{equation}
where $k := \omega / c$ is the wavenumber and $c := 1 / \sqrt{\mu_{0} \epsilon_{0}}$ is the speed of light in vacuum.

Equation (\ref{equation4}) admits a variety of solutions \cite{zhan}. Our objective is to explore the solutions that support an OSS. To this end, we first examine an azimuthal field configuration which is the cylindrical analog of the radial modes of the spherical models studied in \cite{pla-2011,prsa-2012}. This is the content of Section~2. We then construct in Section~3 a set of solutions of the Maxwell equations where the electric field is aligned along the $z$-direction. These include the WGMs as well as a large class of surface modes that support OSSs. In Sections~4 and 5, we give a detailed discussion of three different notions of WGMs. These are the conventional WGMs that are related to the zeros of the Bessel functions $J_\nu$, the WGMs whose energy density has a peak on the surface of the cylinder, and a new class of WGMs that yield spectral singularities. In Section~6, we give a detailed analysis of these WGMs that takes into account the dispersion effects. Finally, in Section~7, we summarize our findings and present our concluding remarks.

\section{Spectral Singularities in the Radial Modes}

Consider a $z$-independent solution of (\ref{equation4}) that has the form
    \be
    \vec{E}= E_{\varphi}\hat{\varphi},
    \label{ansatz1}
    \ee
where $E_{\varphi}$ is a scalar function of $\rho$ and $\varphi$, and a hat on any of the cylindrical coordinates denotes the unit vector along the coordinate in question. Substituting this ansatz in (\ref{equation4}), we find that $E_{\varphi}$ is independent of $\varphi$ and that it satisfies Bessel's equation with the following solution:
    \be
    E_{\varphi}=\left\{
    \begin{aligned}
    &b_1 J_{1} (\fn k\rho) && {\rm for}~\rho<a,\\
    &a_1 H_{1}^{(1)} (k \rho) + a_2 H_{1}^{(2)} (k \rho) && {\rm for}~\rho>a,
    \end{aligned}\right.
    \label{ansatz2}
    \ee
where $J_{1}$ and $H_{1}^{(1,2)}$ are respectively the Bessel and Hankel functions of order 1, and $a_{1,2}$ and $b_1$ are complex coefficients. Inserting (\ref{ansatz1}) in the first equation in (\ref{equation2}) and using (\ref{ansatz2}) to compute $\vec B$, we also find that $\vec B= B_z\hat z$, where
     \be
    B_z=\left\{
    \begin{aligned}
    &-ic^{-1}\fn b_1\tilde{J}_1(\fn k \rho)  && {\rm for}~\rho<a,\\
    &-ic^{-1}\left[a_1\tilde{H}_1^{(1)}(k\rho)+a_2\tilde{H}_1^{(2)}(k\rho)\right] && {\rm for}~\rho>a,
    \end{aligned}\right.
    \label{ansatz3}
    \ee
and for every differentiable function $f$ of a real or complex variable $\rz$,
$\tilde f(\rz):=\left[\frac{d}{d\rz}+\frac{1}{\rz}\right]f(\rz)$.

Having obtained explicit expressions for the electric and magnetic fields inside and outside the gain medium, we use the appropriate boundary conditions for the problem to match them at the boundary of the cylinder \cite{jackson}. These turn out to correspond to the requirement that $E_\varphi$ and $B_z$ be continuous functions at $\rho=a$. In view of (\ref{ansatz2}) and (\ref{ansatz3}), this condition takes the form:
    \bea
    &&b_1 J_{1} (\fn k a) = a_1 H_{1}^{(1)} (k a) + a_2 H_{1}^{(2)}(k a),
    \label{equation8}\\
    &&\fn b_1\tilde{J}_1(\fn k a)=a_1\tilde{H}_1^{(1)}(k a)+a_2\tilde{H}_1^{(2)}(k a).
    \label{equation9}
    \eea
We can use these relations to read off the following expression for the reflection amplitude of the system \cite{pla-2011}.
    \bea
    R := \frac{a_1}{a_2}&=&\left.\frac{J_1(\fn k a)\partial_\rho{H}_1^{(2)}(k\rho)-H_1^{(2)}(k a)\partial_\rho{J}_1(\fn k\rho)}{H_1^{(1)}(k a)\partial_\rho{J}_1(\fn k \rho)-J_1(\fn k a)\partial_\rho{H}_1^{(1)}(k \rho)}\right|_{\rho=a}
    \label{R=az}
    \eea
The spectral singularities are the real and positive values of $k^2$ for which $R$ diverges. They are, therefore, given by the real values of $k$ satisfying
    \be
    H_{1}^{(1)} (k a) \,\partial_\rho J_{1}(k\fn \rho)\Big|_{\rho=a} =
    J_{1}(k\fn a) \, \partial_\rho H_{1}^{(1)}(k\rho)\Big|_{\rho=a}.
    \label{equation11}
    \ee
This is the well-known resonance condition (also called eigenvalue \cite{rowland} or characteristic \cite{oraevsky} equation) that is now required to be satisfied by real values of $k$. In particular, if we define the quality factor $Q$ for the resonant solutions as the ratio of the real part of $k$ to its imaginary part \cite{oraevsky}, we find that the solutions corresponding to a spectral singularity have a divergent quality factor. This is an alternative way of expressing the physical interpretation of a spectral singularity \cite{prl-2009}.

Eq.~(\ref{equation11}) is similar to the one giving the spectral singularities in the radial modes of the spherical gain medium of Ref.~\cite{pla-2011} where the spherical Bessel and Hankel functions appear and the order of these functions is $\sqrt{5}/2$ rather than $1$. Carrying out the same analysis as the one outlined in \cite{pla-2011} and using the same dye laser material (with specifications given in (\ref{specific}) below) we find that OSSs can appear provided that the radius of the cylinder is larger than about $3.28$~mm. This is almost identical with the lower bound we found in \cite{pla-2011} for the radius of the spherical gain medium.

As we pointed out in Section~1, the appearance of such a large lower bound on the size of the gain medium is related to the fact that for the radial modes the length of the optical path along which the wave gets amplified is proportional to the radius of the cylinder. In order to obtain a substantially longer optical path, either we should coat the cylinder with a high-reflectance coating to produce a cylindrical resonator in which the effective optical path is enhanced as a result of multiple internal reflections, or consider field configurations in which the wave circles around the symmetry axes of the cylinder several times before exiting the gain region. The extreme scenario in which this happens is a WGM where the energy density of the wave is concentrated on or near the surface of the cylinder and the Poynting vector is parallel to this surface. This qualitative reasoning requires a quantitative verification that we offer in the following sections.

\section{Spectral Singularities in the Azimuthal Modes}

Consider a transverse electric wave given by $\vec{E} =E_{z} (\rho, \varphi)\,\hat{z}$. Inserting this ansatz in (\ref{equation4}) and seeking for separable solutions corresponding to $E_{z} (\rho, \varphi) = R_{z} (\rho) \Phi_{z} (\varphi)$ give
    \be
    \vec E=\left\{
    \begin{aligned}
    & b_1 J_{\nu} (\fn k\rho)e^{i\nu\varphi}\,\hat z && {\rm for}~\rho<a,\\
    & \left[a_1 H_{\nu}^{(1)} (k \rho) + a_2 H_{\nu}^{(2)} (k \rho)\right]
    e^{i\nu\varphi}\,\hat z~~~~~~ && {\rm for}~\rho>a,
    \end{aligned}\right.
    \label{ansatz12}
    \ee
where $\nu$ is an integer, and $a_1,a_2$, and $b_1$ are complex coefficients possibly depending on $\nu$. Substituting (\ref{ansatz12}) in the first equation in (\ref{equation2}) and solving for $\vec B$, we find
    \be
    \vec B =\left\{
    \begin{aligned}
    & B^{(\rm in)}_\rho\,\hat\rho+B^{(\rm in)}_\varphi\,\hat\varphi
    && {\rm for}~\rho<a,\\
    & B^{(\rm out)}_\rho\,\hat\rho+B^{(\rm out)}_\varphi\,\hat\varphi
    && {\rm for}~\rho>a,
    \end{aligned}\right.
    \label{ansatz22}
    \ee
where
    \bea
    B^{(\rm in)}_\rho&:=&
    b_1\,\nu(\omega\rho)^{-1} J_\nu(\fn k\rho)\, e^{i\nu\varphi},\nn\\
    B^{(\rm in)}_\varphi&:=&
    i b_1\omega^{-1}\partial_\rho J_\nu(\fn k\rho)\,e^{i\nu\varphi},\nn\\
    B^{(\rm out)}_\rho&:=&
    \nu(\omega\rho)^{-1}\left[a_1H_\nu^{(1)}(k\rho)+a_2H_\nu^{(2)}(k\rho)\right]
    e^{i\nu\varphi},\nn\\
    B^{(\rm out)}_\varphi&:=&
    i\omega^{-1}\left[a_1\partial_\rho H_\nu^{(1)}(k\rho)+
    a_2\partial_\rho H_\nu^{(2)}(k\rho)\right]e^{i\nu\varphi}.\nn
    \eea

The coefficients $a_1,a_2$, and $b_1$ are clearly not independent. In order to find the relation among them, we match the values of $\vec E$ and $\vec B$ at the boundary by imposing the appropriate boundary conditions at $\rho=a$, \cite{jackson}. Again, these demand that $\vec E$ and $\vec B$ be continuous functions of $\rho$ at $\rho=a$. More specifically, we have
    \begin{align}
    &b_1 J_{\nu} (\fn k a) = a_1 H_{\nu}^{(1)} (ka) + a_2 H_{\nu}^{(2)} (ka),
    \label{equation29}\\
    &b_1 \partial_\rho J_{\nu} (\fn k \rho)\Big|_{\rho = a}=
    a_1 \partial_\rho H_{\nu}^{(1)}(k \rho)\Big|_{\rho = a}+
    a_2 \partial_\rho H_{\nu}^{(2)}(k\rho)\Big|_{\rho = a}.
    \label{equation30}
    \end{align}
Eliminating $b_1$ in (\ref{equation29}) and (\ref{equation30}), we obtain the following expression for the reflection amplitude.
      \be
      R:= \frac{a_1}{a_2} = \left.\frac{J_{\nu}(\fn k a)
      \partial_\rho H_{\nu}^{(2)}(k \rho) -H_{\nu}^{(2)} (k a)
      \partial_\rho J_{\nu}(\fn k \rho) }{H_{\nu}^{(1)}(k a) \partial_\rho J_{\nu}(\fn k \rho)  - J_{\nu} (\fn k a) \partial_\rho H_{\nu}^{(1)}(k \rho) }\right|_{\rho=a}.
      \label{equation18}
      \ee
In particular, we have an spectral singularity provided that we can satisfy
       \be
       H_{\nu}^{(1)}(k a) \partial_\rho J_{\nu}(\fn k \rho)\Big|_{\rho=a}=
	   J_{\nu} (\fn k a) \partial_\rho H_{\nu}^{(1)}(k \rho)\Big|_{\rho=a}
       \label{equation19}
       \ee
using a real value of $k$. Note that (\ref{equation18}) and (\ref{equation19}) respectively reduce to (\ref{R=az}) and (\ref{equation11}) for $\nu=1$. Equation (\ref{equation19}) is also identical in structure with the equation for the spectral singularities in the radial modes of the spherical model of Ref.~\cite{pla-2011}. The only difference is that in \cite{pla-2011} we have spherical Bessel and Hankel functions of order $\sqrt 5/2$, whereas here we have the usual Bessel and Hankel functions of an arbitrary integer order $\nu$. As we will see below, it is this freedom in the choice of $\nu$ and the fact that it can take arbitrarily large values that allow for realizing spectral singularities in samples with a much smaller radius than those we consider in \cite{pla-2011,prsa-2012}.

\section{Whispering Gallery Modes}

\begin{figure}
    \begin{center}
    \includegraphics[scale=.6]{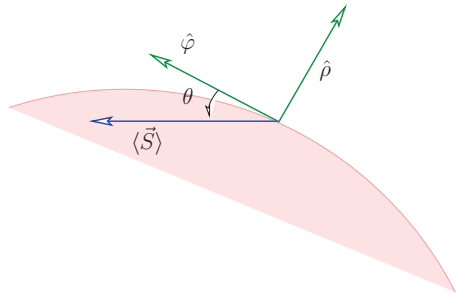}
    \caption{(Color online) The Poynting vector $\br\vec S\kt$ and the angle $\theta$ on the boundary of the cylindrical gain medium. $\hat\rho$ and $\hat\varphi$ are respectively the unit vectors along the radial and azimuthal directions. $\theta$ and the $\rho$-component of $\br\vec S\kt$ have opposite sign.
    \label{fig2}}
    \end{center}
    \end{figure}

Whispering gallery models correspond to field configurations that propagate in a close vicinity of the boundary of a cylindrical or spherical medium. For our cylindrical model, (\ref{ansatz12}) and (\ref{ansatz22}) give an infinite family of exact solutions of the wave equation. In order to see if this family includes whispering gallery models, we examine the corresponding (time-averaged) energy density and Poynting vector. These are respectively given by
     \begin{align*}
     &\br u \kt=\frac{1}{4}\,\RE\left(\vec E\cdot\vec D^* + \vec B\cdot\vec H^*\right),
     &&\br\vec S\kt=\frac{1}{2\mu_0}\,\RE\left(\vec E\times\vec B^*\right).
     \end{align*}
Inserting  (\ref{ansatz12}) and (\ref{ansatz22}) in these equations and carrying out the necessary algebra, we find that for $\rho\leq a$,
    \bea
    \br u\kt &=&\frac{\left|b_1J_\nu(\fn k\rho)\right|^2}{4\mu_0 c^{2} k^{2}}
    \left[k^{2} \,\RE(\fn^2) + \frac{\nu^{2}}{\rho^{2}}+
    \left|\frac{\partial_\rho J_\nu(\fn k\rho)}{J_\nu(\fn k\rho)}\right|^2\right],
    \label{u=}\\
    \br\vec S\kt&=&
    \frac{\left|b_1J_\nu(\fn k\rho)\right|^2}{2\mu_0 c k}\left\{\frac{\nu}{\rho}\,\hat\varphi+
    \IM\left[\frac{\partial_\rho J_\nu(\fn k\rho)}{
    J_\nu(\fn k\rho)}\right]\hat\rho\right\}.
    \label{S=}
    \eea
Let $\theta$ be the angle between $\hat\varphi$ and $\br\vec S\kt$ that is shown in Figure~\ref{fig2}. According to (\ref{S=}) it is given by
    \be
    \theta= -\tan^{-1}\left\{\frac{\rho}{\nu}\,\IM\left[\frac{\partial_\rho J_\nu(\fn k\rho)}{
    J_\nu(\fn k\rho)}\right]\right\}.
    \label{theta}
    \ee
If the cylinder has a real refractive index, $\theta=0$ and our solutions correspond to waves circling around the symmetry axis of the cylinder without any absorption or gain. This is to be expected, because in this case the system does not include an active component. If the cylinder consists of a gain or loss material, then $\theta\neq 0$ and $\br\vec S\kt$ has a radial component. This is an indication that in this case the system emits or absorbs electromagnetic energy.

The solutions we have constructed above correspond to a WGM provided that $\br u\kt$ attains its maximum and and $|\theta|\ll 1$ on or near the surface of the cylinder. For realistic optically active material the imaginary part $\kappa$ of the refractive index is several orders of magnitude smaller than its real part $\eta$. This suggests that we can obtain a rather accurate description of the behavior of $\br u\kt$ and $\theta$ by expanding the terms appearing in (\ref{u=}) and (\ref{theta}) in powers of $\kappa$ and neglecting the second and higher order terms. This gives
	\bea
	\br u\kt&=&\frac{\eta^{2}\left|b_1\right|^2F_+(\zeta)}{4\mu_0 c^{2}}+\mathcal{O}(\kappa^2),
    \label{energy-density-expansion}\\
    \theta&=&\frac{\zeta^2 F_-(\zeta)\,\kappa}{\eta\,\nu J_\nu(\zeta)^2}+\mathcal{O}(\kappa^2),
    \label{theta-2}
    \eea
where $\zeta:=\eta k\rho$, $F_\pm(\zeta):=
    J_\nu^{'}(\zeta)^2+\left(1\pm\frac{\nu^2}{\zeta^2}\right)J_\nu(\zeta)^2$,
a prime denotes the derivative of the corresponding function, and $\mathcal{O} (\kappa^\ell)$ stands for terms of order $\ell$ and higher in $\kappa$.

\begin{figure}
    \begin{center}
	\includegraphics[scale=.5]{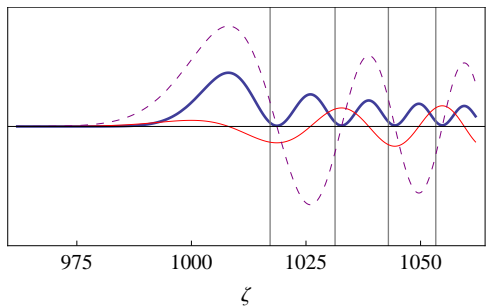}\\
    \caption{(Color online) Graphs of $F_+$ (thick blue curve), $J_\nu$ (dashed purple curve), and $J'_\nu$ (thin red curve) as functions of $\zeta$ for $\nu=1000$. The peaks of $F_+$ (and therefore $\br u\kt$) are attained at the positive zeros of $J'_\nu$. The highest pick occurs at $\zeta= j'_{\nu,1}\approx 1008.1$, which is the smallest positive zero of $J'_\nu$. The vertical gray lines give the $\zeta$ values corresponding to a spectral singularity for $\eta=1.479$. The smallest of these is about 1017.2. This is larger than $j'_{\nu,1}$ and smaller than the smallest positive zero of $J_\nu$, namely $j_{\nu,1}\approx 1018.7$.}
    \label{fig3}
	\end{center}
	\end{figure}

It turns out that $F_+$ has an infinity of positive maximum points.  The smallest of these that yields the highest peak (global maxium) of $F_+$ is given by
	\be
	\zeta\approx \nu\left[1+ 0.81\, \nu^{-\frac{2}{3}}+\cO(\nu^{-\frac{4}{3}})\right].
	\label{j-zero-1}
	\ee
This is slightly larger than $\nu$,  \cite{abramowitz}. Demanding that this maximum point occurs at the boundary of the cylinder, $\rho=a$, so that $\br u\kt$ has no maxima inside the cylinder, we find for physically relevant situations that $\zeta=\eta\,k a\gg 1$. Therefore $\zeta$ is a maximum point of $F_+$ provided that $\zeta\gtrsim \nu\gg 1$. For this range of values of $\zeta$ and $\nu$, $(1 +\nu^{2}/\zeta^{2})J_\nu(\zeta)^2$ can take much larger values than $J'_\nu(\zeta)^2$. This suggests that the maxima of $F_+$ (and therefore $\br u\kt$) coincide with those of $J_\nu(\zeta)^2$, i.e., they are the zeros of $J'_\nu(\zeta)$. This observation can be easily verified graphically, as shown in Figure~\ref{fig3}.

Next, we obtain a crude estimate for the magnitude of the angle $\theta$ at the smallest zero of $J'_\nu$ where $\br u\kt$ attains its global maximum value. In view of (\ref{theta-2}) and (\ref{j-zero-1}) and the fact that $\nu\gg 1$, we find
	\be
	\theta\approx \frac{1.6\nu^{\frac{1}{3}}}{\eta}
	\left[1+1.6\nu^ {-\frac{1}{3}}+\cO(\nu^{-1})\right]\kappa+\cO(\kappa^2).
	\label{theta-3}
	\ee
Typically $\kappa$ and $\nu$ are at most of the order of $10^{-3}$ and $10^3$, which combined with the fact that $\eta>1$ correspond to $\theta$ values of at most of the order of $10^{-2}$.

In light of (\ref{theta-2}) and the fact that the smallest zero of $J'_\nu$ is slightly larger than $\nu$, we see that up to terms of order $\kappa^2$, that we can safely neglect, $\theta$ has the same sign as $\kappa$. This means that whenever the cylinder is filled with a gain material, in which case $\kappa<0$, $\theta$ is negative and the radial component of the Poynting vector points away from the cylinder. This confirms the expectation that in this case there is a flux of energy through the boundary of the cylinder that is directed away from it. Conversely for a lossy material $\kappa$ and consequently $\theta$ are positive, there is an energy flux pointing toward the symmetry axis of the cylinder, and the system absorbs energy.

The above observations show that our solutions become WGMs with wavenumber $k$ provided that $J_\nu^{'}(\eta k a)\approx 0$. This is actually not quite the same as the standard description of WGMs given in the literature \cite{oraevsky} where they are associated with the zeros of $J_\nu$ rather than those of $J'_\nu$. The reasoning used in the standard description of the electromagnetic WGMs has its root in Lord Rayleigh's seminal works on acoustic WGM and is based on the following properties of $J_\nu$, \cite{abramowitz}.
	\begin{enumerate}
	\item The $\rho$-dependence of the (electric) field inside the cylinder (or a sphere) is essentially determined by $J_\nu(\eta k\rho)$.
	\item For $\nu\gg 1$, which is the case for almost all practical purposes,  increasing the value of $\zeta$ starting from zero has the following effect on $J_\nu(\zeta)$. $J_\nu(\zeta)$ takes negligibly small values for $0\leq \zeta<\nu/2$ (with $J_\nu(0)=0$). It then attains its largest value at the first zero of $J'_\nu(\zeta)$, namely
    \be
    j'_{\nu,1}=\nu\left[1+ 0.809\, \nu^{-\frac{2}{3}}+\cO(\nu^{-\frac{4}{3}})\right],
    \label{j-prim-zero}
    \ee
which is slightly larger than $\nu$. As $\zeta$ exceeds $j'_{\nu,1}$,  $J_\nu(\zeta)$ decreases until it hits its first zero,
    \be
    j_{\nu,1}=\nu\left[1+ 1.856\, \nu^{-\frac{2}{3}}+\cO(\nu^{-\frac{4}{3}})\right],
    \label{j-zero}
    \ee
which is slightly larger than $j'_{\nu,1}$. Increasing $\zeta$ further makes $J_\nu(\zeta)$ oscillate between positive and negative values with a diminishing amplitude. Figure~\ref{fig3} provides a graphical demonstration of this behavior for $\nu=1000$.
	\end{enumerate}
According to Lord Rayleigh a WGM corresponds to the situation that the wave is ``pressed down'' to the surface \cite{oraevsky}. This requirement is fulfilled provided that we take $\eta k a=j_{\nu,1}$. This is because, in this case the largest peak of the energy density resides inside the cylinder and in a close vicinity of its boundary. In contrast, the condition $\eta k a=j_{\nu,1}'$ implies that this peak lies on the boundary of the cylinder. This is also consistent with Lord Rayleigh's description of a WGM. Therefore, $\eta k a=j_{\nu,1}$ and $\eta k a=j_{\nu,1}'$ correspond to different types of WGMs that we respectively denote by WGM and WGM$'$. The same holds for higher order WGMs that are associated with larger zeros of $J_\nu$ and $J_\nu'$. These are labeled by the azimuthal mode number $\nu$ and a radial mode number $q$ that counts the zeros of $J_\nu$ (or $J_\nu'$) \cite{oraevsky}. Because typically $\nu$ takes very large numbers, WGM and WGM$'$ are quite closely located and as a result have similar physical properties.

\section{Singular Gallery Modes}

In Section~3 we showed that the field configurations (\ref{ansatz12}) and (\ref{ansatz22}) support an OSS provided that they satisfy (\ref{equation19}). In Section~4 we derived conditions under which we could identify them with WGMs. In this section, we examine WGMs that support an OSS. In order to gain an understanding of the problem, we offer an analytic treatment of (\ref{equation19}) based on the first-order perturbation theory in which we ignore the second and higher order terms in powers of $\kappa$ and $\nu^{-1}$.

First, we express (\ref{equation19}) as
	\be
	\frac{\fn\,J'_\nu(\fn x)}{J_\nu(\fn x)}=\frac{H_\nu^{(1)\prime}(x)}{H_\nu^{(1)}(x)},
	\label{B2-1}
	\ee
where $x:=ka$. We then recall that $\fn\,x=(\eta+i\kappa)x=\zeta+i\kappa x$ and expand the left-hand side of (\ref{B2-1}) in powers of $\kappa$. This gives
	\be
	\frac{\fn\,J'_\nu(\fn x)}{J_\nu(\fn x)}= \frac{\eta J'_\nu(\zeta)}{J_\nu(\zeta)}
	-i\zeta\left[\frac{J^{'}_\nu(\zeta)^2}{J_\nu(\zeta)^2}+
	1-\frac{\nu^2}{\zeta^2}\right]\kappa+\cO(\kappa^2).
	\label{B2-2}
	\ee

Next, we recall that for a WGM, $\zeta>\nu\gg 1$. This suggests that in our perturbative expansion of the right-hand side of (\ref{B2-2}) we use the Debye's asymptotic expansions \cite{abramowitz}:
	\be
    J_\nu(\zeta)=\sqrt{\frac{2}{\pi\nu\tan\alpha}}
    \left[\cos\phi+\cO(\nu^{-1})\right],~~~~~ J'_\nu(\zeta)=\sqrt{\frac{2}{\pi\nu\tan\alpha}}
    \left[\sin\phi+\cO(\nu^{-1})\right],
	\label{B2-3}
	\ee
where $\alpha:=\cos^{-1}(\nu/\zeta)\in(0,\frac{\pi}{2})$ and $\phi:=\nu(\tan\alpha-\alpha)-\frac{\pi}{4}$. Furthermore, for the typical laser material that we are interested, $\eta>1$ and $|\zeta-\nu|\ll \nu$ so that $x<\nu$. This in turn shows that we can use the Debye's asymptotic expansions \cite{abramowitz}:
	\be
	H_\nu^{(1)}(x)=\frac{e^\psi-2i e^{-\psi}+\cO(\nu^{-1})}{\sqrt{2\pi\nu\tanh\beta}},~~~~~
	H_\nu^{(1)\prime}(x)=\sqrt{\frac{\sinh(2\beta)}{4\pi\nu}}
	\left[e^\psi+2i e^{-\psi}+\cO(\nu^{-1})\right],
	\label{B2-4}
	\ee
where $\beta:=\cosh^{-1}(\nu/x)\in\R^+$ and $\psi:=\nu(\tanh\beta-\beta)$.
Substituting (\ref{B2-3}) in (\ref{B2-2}) and using the resulting expression and (\ref{B2-4}) in (\ref{B2-1}), we find a complex equation whose real and imaginary parts give
	\begin{align}
	&\left(\frac{1-4e^{-4\psi}}{1 + 4e^{-4\psi}}\right) \sinh\beta +
	\eta\sin\alpha\tan\phi \approx 0,
	\label{spec-sing1}\\
	&\left(\frac{4e^{-2\psi}}{1 + 4e^{-4\psi}}\right) \sinh\beta+
	\left(\frac{x\,\eta\sin^{2}\alpha}{\cos^{2}\phi}\right)\kappa \approx 0.
	\label{spec-sing2}
	\end{align}
Here $\approx$ stands for approximate equalities that involve ignoring $\cO(\kappa^2)$ and $\cO(\eta^{-1})$ in (\ref{B2-2}) -- (\ref{B2-4}). Moreover using (\ref{B2-3}) in (\ref{theta-2}), we have
    \be
    \theta\approx\frac{\zeta^2\kappa}{\eta\nu}\left(\sec^2\phi-\frac{\nu^2}{\zeta^2}\right).
    \label{spec-sing3}
	\ee

Equations (\ref{spec-sing1}) and (\ref{spec-sing2}) describe OSSs in the azimuthal modes fulfilling the condition
    \be
    \zeta>\nu>x\gg 1\gg|\kappa|.
    \label{condi1}
    \ee
In particular, (\ref{spec-sing2}) implies that no spectral singularity can exist for a passive or lossy medium where $\kappa\geq 0$, \cite{pra-2009}.

For a WGM$'$, where $J'_\nu(\zeta)\approx 0$, (\ref{B2-3}) together with the fact that $0<\alpha\ll \pi/2$ give $\sin\phi\approx 0$ and consequently $\tan\phi\approx 0$. Inserting this relation in (\ref{spec-sing1}) and making use of the fact that $\sinh\beta>0$, we obtain
	\be
	\psi\approx \frac{\ln 2}{2}\approx 0.347.
	\label{nogo}
	\ee
Note, however, that for $\beta\geq 0$, $\psi$ is a decreasing function of $\beta$. Because $\psi$ vanishes for $\beta=0$, we have $\psi\leq 0$ for all $\beta\geq 0$. This shows that we can never satisfy (\ref{nogo}). Therefore, WGM$'$s do not support spectral singularities.

The argument we used to establish the impossibility of producing an OSS in a WGM$'$ relies only on the conditions (\ref{condi1}) and $J'_\nu(\zeta)\approx 0$. Therefore, it holds also for higher order WGM$'$ whose energy density $\br u\kt$ has peaks inside as well as on the boundary of the cylinder. This argument does not, however, exclude the possibility of realizing spectral singularities for a class of azimuthal modes such that $\rho=a$ is not a maximum of $\br u\kt$ but the Poynting vector is still along the tangential direction to the surface of the cylinder, i.e., $|\theta|\ll 1$. The conventional WGMs that belong to this class turn out not to support OSS either. This can be justified as follows. Consider the WGM corresponding to the first zero of $J_\nu$, i.e., $\zeta=j_{\nu,1}$. According to (\ref{B2-3}), this implies that $\cos\phi\approx 0$. This is consistent with (\ref{spec-sing1}) only if the absolute value of the first term on the left-hand side of (\ref{spec-sing1}) takes extremely large values. It is not difficult to see that according to (\ref{condi1}) this quantity is bounded from above. Therefore, this WGM does not support an OSS. With further care one can also show that the same conclusion holds for WGMs with larger radial mode numbers.

The inconsistency between the equations determining OSSs and those defining WGMs and WGM$'$s suggests that we impose the former equations on field configurations with large $\nu$ values directly. This defines a new class of surface modes that are consistent with Lord Rayleigh's qualitative description of WGMs and yield OSSs. We call them ``singular gallery modes (SGMs).''

The vertical gray lines in Figure~\ref{fig3} correspond to the $\zeta$ values at which a sample with $\eta=1.479$ displays an OSS for $\nu=1000$. They represent SGMs. As shown in this figure for the first four WGMs, fixing the value of $\nu$ we find a single SGM for each WGM (zero of $j_\nu$). This suggests labeling the SGMs by the same azimuthal mode number $\nu$ and the radial mode number $q$ that we use to label WGMs and WGM$'$s. We  respectively use $\lambda_{\nu,{q}}$, $\kappa_{\nu,{q}}$, $g_{\nu,{q}}$, and $\theta_{\nu,{q}}$ to denote the wavelength $\lambda$, imaginary part of the complex refractive index $\kappa$, the gain coefficient $g=-4\pi\kappa/\lambda$, and the angle $\theta$ for the corresponding OSS. Table~\ref{table1} gives the values of these quantities for a sample of radius $a=75\,\mu{\rm m}$.
    \begin{table}
    \begin{center}
    \begin{tabular}{|c|c|c|c|c|c|}
    \hline
    ${q}$ & $\zeta$ & $\lambda_{\nu,{q}}$ (nm) & $\kappa_{\nu,{q}}$ & $g_{\nu,{q}}\:({\rm cm}^{-1})$ & $\theta_{\nu,{q}}$ \\
    \hline
    1 & 1017.171 & 685.196 & $-4.272 \times 10^{-172}$ & $7.836 \times 10^{-167}$ & $-4.558 \times 10^{-168}$\\
    2 & 1031.343 & 675.781 & $-1.917 \times 10^{-163}$ & $3.565 \times 10^{-158}$ & $-1.121 \times 10^{-159}$\\
    3 & 1042.981 & 668.240 & $-1.649 \times 10^{-156}$ & $3.100 \times 10^{-151}$ & $-7.040 \times 10^{-153}$\\
    4 & 1053.324 & 661.679 & $-1.781 \times 10^{-150}$ & $3.384 \times 10^{-145}$ & $-6.148 \times 10^{-147}$\\
    $\vdots$ &$\vdots$&$\vdots$&$\vdots$&$\vdots$&$\vdots$\\
    80 & 1462.126 & 476.642& $-1.938 \times 10^{-5}$ & $5.110$ & $-1.546 \times 10^{-2}$\\
    81 & 1466.506 & 475.254& $-3.733 \times 10^{-5}$ & $9.871$ & $-2.980 \times 10^{-2}$\\
    82 & 1470.756 & 473.880& $-5.759 \times 10^{-5}$ & $15.273$ & $-4.595 \times 10^{-2}$\\
    83 & 1474.984 & 472.522& $-6.171 \times 10^{-5}$ & $16.411$ & $-4.923 \times 10^{-2}$\\
    \hline
    \end{tabular}
    \vspace{6pt}
    \caption{The values of $\lambda_{\nu,{q}}$, $\kappa_{\nu,{q}}$, $g_{\nu,{q}}$, and $\theta_{\nu,{q}}$ for spectral singularity of the lasing gallery modes with $\nu=1000$ for a sample of radius $a=75\:\mu{\rm m}$ and real part of complex refractive index $\eta=1.479$. The radial mode number ${q}$ takes values between 1 and 83.}
    \label{table1}
    \end{center}
    \end{table}
Because the set of zeros of $j_\nu$ does not have an accumulation point, ${q}$ has an upper bound $q_{\rm max}$. This means that for each $\nu$ there are finitely many SGMs and OSSs. Furthermore, because the larger values of $\zeta$ correspond to larger values of $x$ and therefore smaller values of the wavelength, $\lambda_{\nu,{q}}$ is a decreasing function of ${q}$. Another interesting result is that $-\kappa_{\nu,{q}}$ and $-\theta_{\nu,{q}}$ are increasing functions of ${q}$ and take extremely small values. The same is true for $g_{\nu,{q}}$ except for ${q}\geq 75$. For $\eta=1.479$, $a=75\,\mu{\rm m}$, and $\nu=1000$, we have ${q}=1,2,\cdots,83$,
    \begin{align*}
    & 472.521<\lambda_{\nu,{q}}/{\rm nm}<685.197, &&
    10^{-172}<-\kappa_{\nu,{q}}<10^{-4},\\
    &10^{-166}<g_{\nu,{q}}/{\rm cm}<17, &&
    10^{-168}<-\theta_{\nu,{q}}<0.05.
    \end{align*}
The small values of $\kappa$ explains the extremely good agreement between the perturbative results obtained using (\ref{spec-sing1}) -- (\ref{spec-sing3}) and the exact (numerical) results that employ (\ref{B2-1}) and (\ref{theta-2}).

Table~\ref{table2} gives the number of radial modes $q_{\rm max}$, the minimum and maximum values of the wavelength, $\lambda_{\rm min}:=\lambda_{\nu,q_{\rm max}}$ and $\lambda_{\rm max}:=\lambda_{\nu,1}$, and the smallest gain coefficient for SGMs with azimuthal mode numbers $\nu=850,900,\cdots,1150$. The interval $[\lambda_{\rm min},\lambda_{\rm max}]$ is the spectral range for these SGMs. As we increase $\nu$ it shrinks slightly and gets red-shifted while $q_{\rm max}$ increases and $g_{\rm min}$ decreases monotonically.
    \begin{table}
    \begin{center}
    \begin{tabular}{|c|c|c|c|c|}
    \hline
    $\nu$ & $q_{\rm max}$ & $\lambda_{\rm min}$ (nm) & $\lambda_{\rm max}$ (nm) &
    $g_{\rm min}\,({\rm cm}^{-1})$ \\
    \hline
    850 & 70 & 555.275 & 804.615 & $1.536\times 10^{-140}$\\
    900 & 74 & 524.647 & 760.428 & $2.693\times 10^{-149}$\\
    950 & 78 & 497.222 & 720.851 & $4.632\times 10^{-158}$\\
    1000 & 83 & 472.522 & 685.196 & $7.836 \times 10^{-167}$\\
    1050 & 87 & 448.938 & 652.909 & $1.305 \times 10^{-175}$\\
    1100 & 92 & 428.703 & 623.533 & $2.144 \times 10^{-184}$\\
    1150 & 95 & 410.214 & 596.691 & $3.477 \times 10^{-193}$\\
    \hline
    \end{tabular}
    \vspace{6pt}
    \caption{Values of $q_{\rm max}$, $\lambda_{\rm min}$, $\lambda_{\rm max}$, and $g_{\rm min}$ for various $\nu$, $\eta=1.479$, and $a=75\,\mu{\rm m}$.}
    \label{table2}
    \end{center}
    \end{table}
This behavior of $g_{\rm min}$ and the fact that it takes extremely small values show that unless we restrict the spectral range of interest, the system supports OSSs and begins lasing for any positive gain coefficient. Consequently, in principle, the system lacks a lasing threshold. This is actually an expected result, for the length of the optical path along which the surface waves get amplified has no upper bound. Our numerical calculations show, however, that the OSS requiring lower gain coefficients have a wavelength that is more sensitive to fluctuations.

\section{Characterization of Singular Gallery Modes}

Equations (\ref{spec-sing1}) and (\ref{spec-sing2}) that describe OSSs involve four real parameters, namely $x, \eta, \kappa$ and $\nu$. Among these $x$ and $\kappa$ depend on the radius $a$, the wavelength $\lambda$, and the gain coefficient $g$ of the medium according to \cite{silfvast}
    \be
    x=ak=\frac{2\pi a}{\lambda},~~~~~~~~~~~\kappa=-\frac{\lambda g}{4\pi}.
    \label{x-kappa=}
    \ee
In the preceding section we used (\ref{spec-sing1}) and (\ref{spec-sing2}) to describe typical examples of optical spectral singularities for a sample whose refractive index was assumed not to depend on the wavelength. In this section we give a more detailed study of the SGMs that take into account the dispersion effects.

Consider an optically active medium that is obtained by doping a host medium of refraction index $n_0$ and modeled by a two-level atomic system with lower and upper level population densities $N_l$ and $N_u$, resonance frequency $\omega_0$, damping coefficient $\gamma$, and the dispersion relation \cite{pra-2011a}:
    \be
    \fn^2= n_0^2-
    \frac{\hat\omega_p^2}{\hat\omega^2-1+i\hat\gamma\,\hat\omega},
    \label{epsilon}
    \ee
where $\hat\omega:=\omega/\omega_0$, $\hat\gamma:=\gamma/\omega_0$,
$\hat\omega_p:=(N_l-N_u)e^2/(m_e\varepsilon_0\omega_0^2)$, $e$ is
electron's charge, and $m_e$ is its mass. We can express
$\hat\omega_p^2$ in terms of the imaginary part $\kappa_0$ of $\fn$
at the resonance wavelength $\lambda_0:=2\pi c/\omega_0$ according
to $\hat\omega_p^2=2n_0\hat\gamma\kappa_0+\cO(\kappa_0^2)$, \cite{pra-2011a}.
Inserting this relation in (\ref{epsilon}), using
$\fn=\eta+i\kappa$, and neglecting $\cO(\kappa_0^2)$, we obtain \cite{pla-2011}
    \begin{align}
    &\eta\approx n_0+\kappa_0f_1(\hat\omega),
    &&\kappa\approx\kappa_0f_2(\hat\omega),
    \label{eqz301}
    \end{align}
where $f_1(\hat\omega):= \hat\gamma(1-\hat\omega^2)/[(1-\hat\omega^2)^2+
    \hat\gamma^2\hat\omega^2]$ and $f_2(\hat\omega):= \hat\gamma^2\hat\omega/[(1-\hat\omega^2)^2+
    \hat\gamma^2\hat\omega^2]$.
In view of (\ref{x-kappa=}), we also have $\kappa_0=-\lambda_{0}g_0/(4\pi)$. Substituting this equation in (\ref{eqz301}) and using the resulting relations together with (\ref{x-kappa=}) in (\ref{spec-sing1}) and (\ref{spec-sing2}) we can determine the $\lambda$ and $g_{0}$ values for the spectral singularities that identify SGMs.

For definiteness we consider a cylinder of radius $a = 75\,\mu{\rm m}$ filled with
a Rose Bengal-DMSO (Dimethyl sulfoxide) solution with the following
characteristics \cite{silfvast,Rose}.
    \begin{align}
    &n_0=1.479, &&\lambda_0=549\,{\rm nm}, &&
    \hat\gamma=0.062, &&g_0\leq 5\,{\rm cm}^{-1}.
    \label{specific}
    \end{align}
The following is a summary of our numerical investigation of the properties of the OSSs for the gain medium (\ref{specific}). We obtained them using the exact equation for spectral singularities (\ref{B2-1}) and the dispersion relations (\ref{eqz301}).

\begin{figure}
    \begin{center}
    \includegraphics[scale=.43]{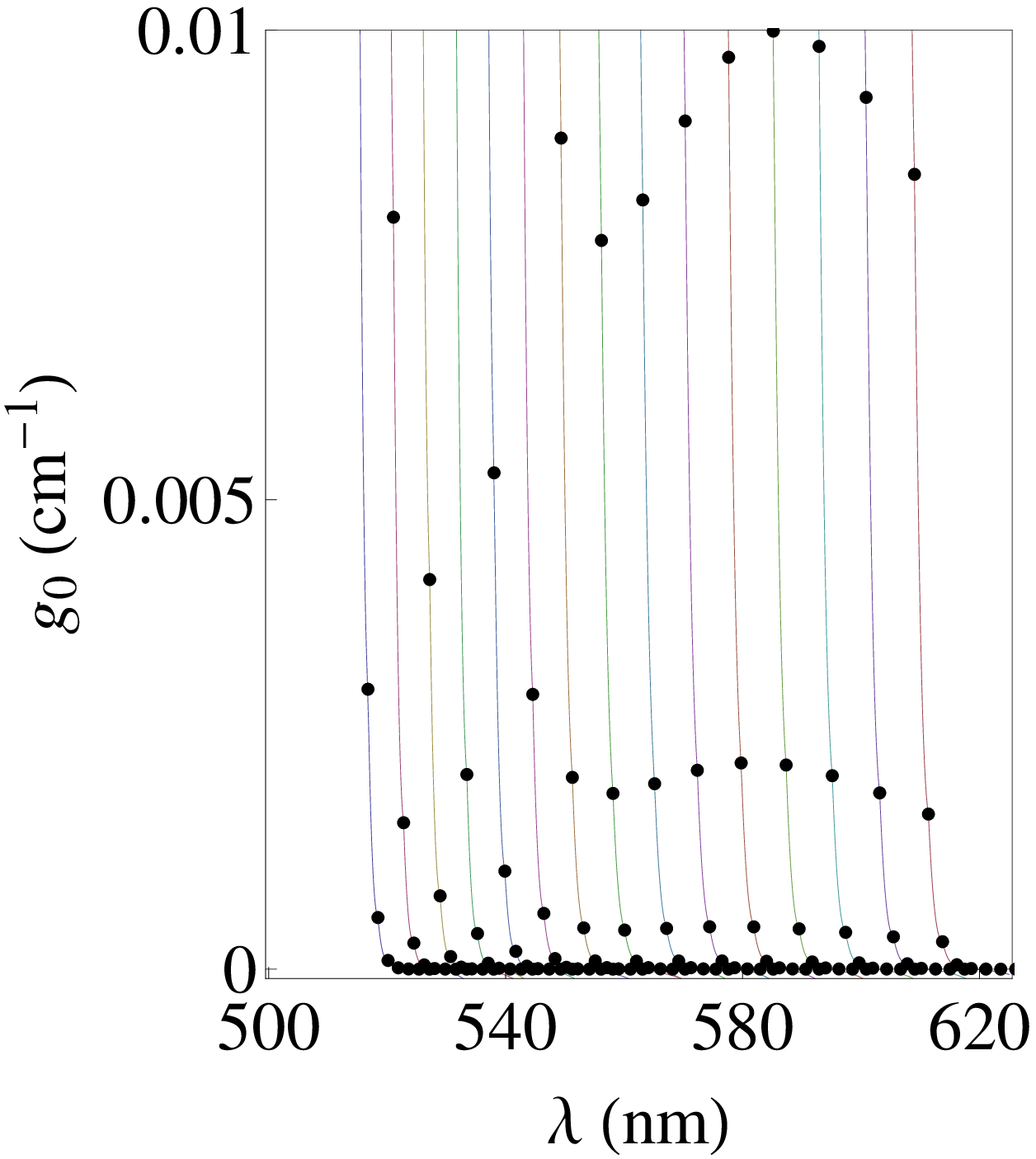}~
	\includegraphics[scale=.43]{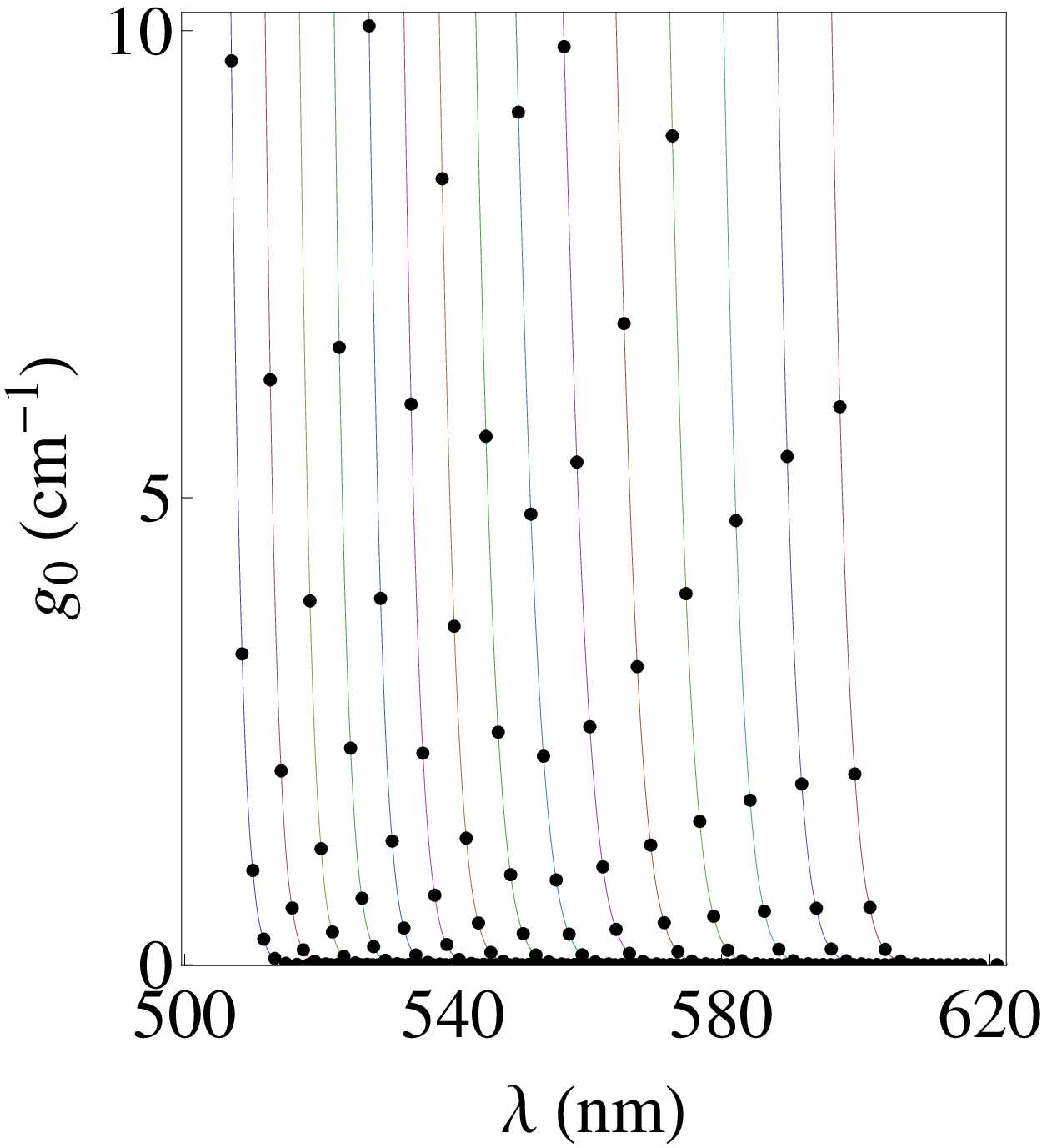}~
	\includegraphics[scale=.43]{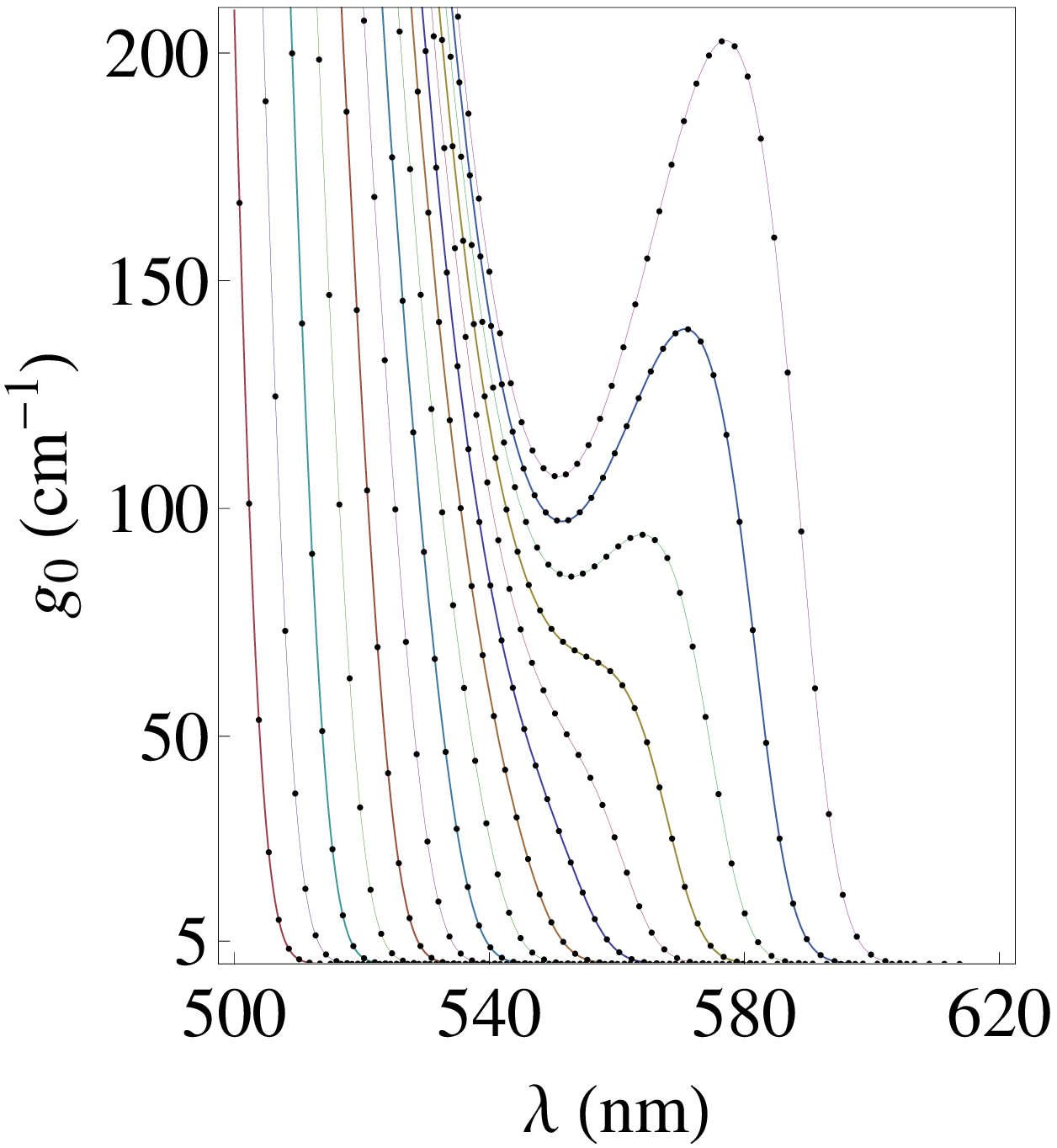}\\
    \caption{(Color online) Spectral singularities associated with the SGMs of a cylindrical sample of radius $75\,\mu{\rm m}$ consisting of the Rose Bengal-DMSO solution (\ref{specific}). Each curve corresponds to a particular value of $\nu$ that ranges from 805 (the rightmost curve) to 945 (the leftmost curve) in increments of 10. The displayed dots represent the spectral singularities.}
    \label{fig4}
    \end{center}
    \end{figure}

Figure~\ref{fig4} shows the location of the OSSs in the $\lambda$-$g_0$ plane for various choices of $\nu$.
The presence of OSSs for extremely small gain coefficients confirms our expectation that surface waves can support OSSs for very small values of the radius.

\begin{figure}
    \begin{center}
    \includegraphics[scale=.40]{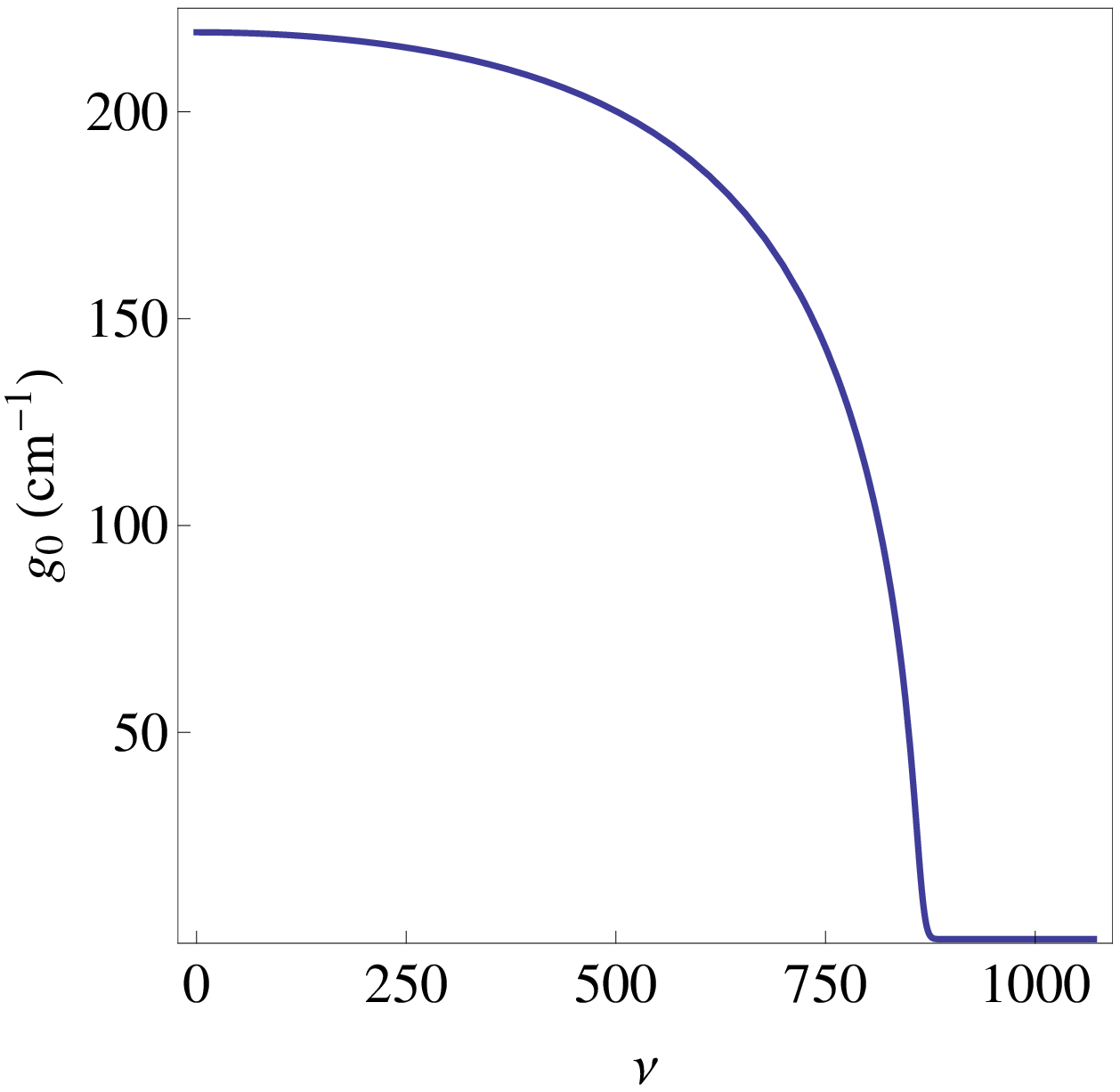}~~~
    \includegraphics[scale=.40]{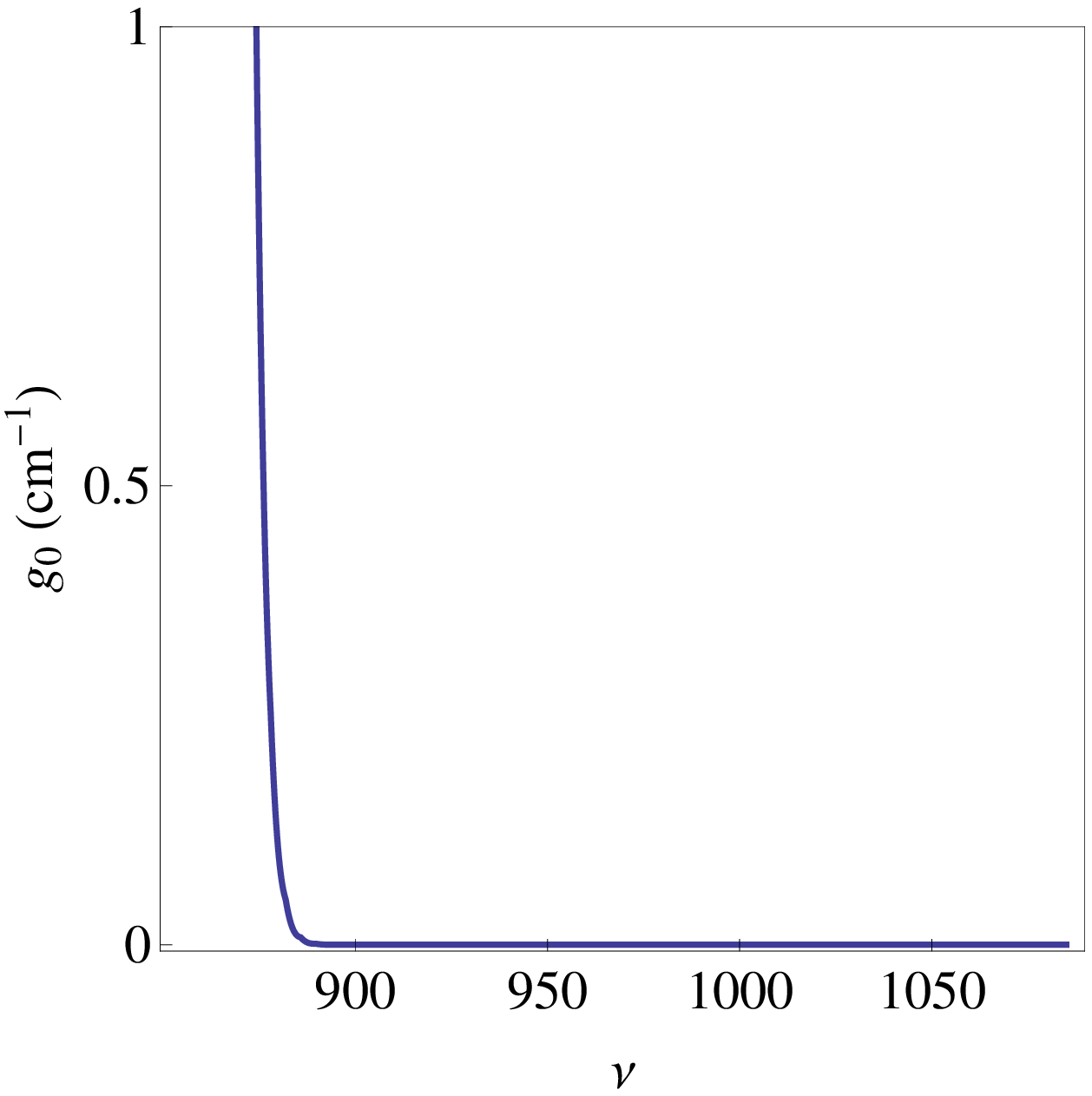}
    \caption{(Color online) Graph of the gain coefficient $g_{0}$ as a function of $\nu$ for an OSS whose wavelength is closest to the resonance wavelength, $\lambda\approx 549~{\rm nm}$. The physical parameters of the gain medium is given by (\ref{specific}) and $a = 75\,\mu{\rm m}$. As $\nu$ tends to zero, $g_0$ approaches to an asymptotic value of about $219\,{\rm cm}^{-1}$. There is a sharp drop of $g_0$ for $\nu\approx 880$.}
    \label{fig5}
    \end{center}
    \end{figure}

Figure~\ref{fig5} shows the behavior of the gain coefficient necessary for generating an OSS with resonance wavelength $\lambda=549~{\rm nm}$ for different values of $\nu$.
The larger $\nu$ is, the smaller the required gain becomes. This confirms the perturbative results we have listed in Table~\ref{table2}.

As we explained in Section~2, the quality factor $Q$ for a field configuration that gives rise to an OSS diverges. In reality the physical parameters that correspond to the emergence of an OSS can be computed with finite precision. The $Q$-factor computed in this way is expected to be finite but very large. If we use more precise values for the physical parameters of the OSS,
we obtain larger $Q$-factors. For example suppose that we fix $\nu$ and $a$, and determine the gain coefficient $g_0$ required for a SGM to an accuracy of $10^{-3}\,{\rm cm}^{-1}$. If we use this value of $g_0$ in the exact equation for OSSs, i.e., (\ref{B2-1}), and solve this equation for the wavenumber $k$, we find a complex value with very small imaginary part. We can use this data to compute the quality factor for the SGM. Table~\ref{table3} gives the result of this calculation for the sample~(\ref{specific}) with $a=75\:\mu{\rm m}$, $\nu=1000$ and $q=65,66,\cdots, 71$. The fact that the $Q$-factor takes very large values is particularly remarkable, for we have examined SGMs with very large radial mode number where the quality factor of the corresponding conventional passive WGM is much smaller \cite{oraevsky}.
    \begin{table}
    \begin{center}
    \begin{tabular}{|c|c|c|c|c|c|}
    \hline
    ${q}$ & $\zeta$ & $\lambda_{\nu,{q}}$ (nm) & $g_{\nu,{q}}\:({\rm cm}^{-1})$ & $Q$ & $Q_0$ \\
    \hline
    65& $1382.989 $ & $503.954$ & $1.190 \times 10^{-11}$ & $1.048\times 10^{20}$ & $6.738\times 10^{16}$\\
    66& $1387.515 $ & $502.310$ & $1.220 \times 10^{-10}$ & $3.103\times 10^{19}$ & $7.011\times 10^{15}$\\
    67 &$1392.024 $ & $500.683$ & $1.165 \times 10^{-9}$  & $2.780\times 10^{18}$ & $7.884\times 10^{14}$\\
    68 & $1395.518$ & $499.072$ & $1.033 \times 10^{-8}$  & $2.932\times 10^{18}$ & $9.490\times 10^{13}$\\
    69 & $1400.996$ & $497.476$ & $8.514 \times 10^{-8}$  & $3.560\times 10^{17}$ & $4.901\times 10^{13}$\\
    70 & $1405.459$ & $495.897$ & $6.516 \times 10^{-7}$  & $4.030\times 10^{16}$ & $6.810\times 10^{12}$\\
    71 & $1409.908$ & $494.332$ & $4.627 \times 10^{-6}$  & $1.504\times 10^{15}$ & $1.018\times 10^{12}$\\
    \hline
    \end{tabular}
    \vspace{6pt}
    \caption{The values of $\zeta$, $\lambda_{\nu,{q}}$, $g_{\nu,{q}}$, and $Q$ for spectral singularity of the SGMs with $\nu=1000$ for a sample with specifications (\ref{specific}), $a=75\:\mu{\rm m}$, $\nu=1000$, and ${q}$ taking values between 65 and 71. $Q_0$ is the value of the quality factor for the corresponding passive mode where $\kappa=0$.}
    \label{table3}
    \end{center}
    \end{table}
	
\begin{figure}
	\begin{center}
    \includegraphics[scale=.40]{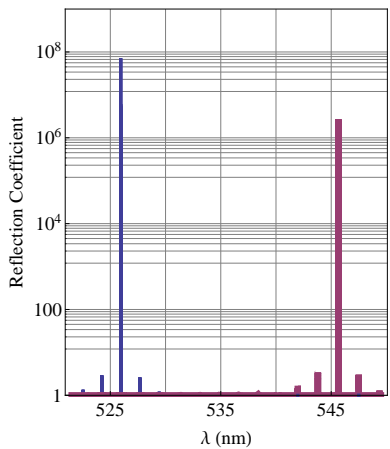}
	\caption{(Color online)  Logarithmic plots of the reflection coefficient $|R|^2$ as a function of $\lambda$ for $a=75\:\mu{\rm m}$, $g_0= 0.132\:{\rm cm}^{-1}$, and $\nu= 920$ (the blue thin curve with the highest pick at $\lambda\approx 526\:{\rm nm}$) and $\nu=885$ (the purple thick curve with the highest pick at $\lambda\approx 546\:{\rm nm}$).}
    \label{fig6}
    \end{center}
    \end{figure}

Because $\nu$ can take arbitrary large values and for each value of $\nu$ there are many SGMs, it is tempting to see if we can generate OSSs with different wavelengths using the same amount of gain. Figure~\ref{fig6} shows the graph of the reflection coefficient $|R|^2$ as a function of $\lambda$ for a situation in which this scenario holds effectively.
The peaks correspond to a pair of OSS that arise for $g_0\approx\tilde g_0:=0.132\,{\rm cm}^{-1}$. They have the following $\nu$, $g_0$, and $\lambda$ values.
	\bea
	&&\nu=920,~~~~~g_0=0.131968\:{\rm cm}^{-1},~~~~~\lambda = 525.945 537\:{\rm nm},\nn\\
    &&\nu=885,~~~~~g_0=0.132161\:{\rm cm}^{-1},~~~~~\lambda = 545.633 973\:{\rm nm}.\nn	
	\eea
The peak on left of Figure~\ref{fig6} has a larger height than the one on the right, simply because the exact value of $g_0$ for the former is closer to $\tilde g_0$ than that of the latter. Our numerical studies show that using more and more precise values for $g_0$ and $\lambda$ for plotting the graph of $|R|^2$ as a function $\lambda$ gives larger and larger values for the hight of the peaks. Therefore their width is zero. This is a clear indication that they represent spectral singularities \cite{prl-2009}.

\section{Discussion and Conclusion}

In this article we have addressed the problem of exploring optical spectral singularities for the radial and azimuthal modes of a cylindrical gain medium. We showed that the conventional whispering gallery modes and those corresponding to situations where the energy density has a peak on the surface of the cylinder do not support spectral singularities. The condition that a surface wave does give rise to a spectral singularity defines a new class of whispering gallery modes. These, by definition, have a divergent quality factor, and as a result we call them singular gallery modes.

Our treatment of whispering gallery modes differs from the standard approach \cite{lam,oraevsky,WGM} that makes use of the uniform asymptotic expansion for the Bessel function $J_\nu$ and its derivative \cite{abramowitz}. The latter yields asymptotic (large-$\nu$) series involving the Airy function ${\rm Ai}$ and its derivative for $J(\nu + t\nu^{1/3})$ and $J'(\nu + t\nu^{1/3})$ where $t$ is a real and positive parameter that is of zeroth order in $\nu$, $t=\cO(\nu^0)$. In our notation, use of this expansion amounts to setting $\eta x=\nu + t\nu^{1/3}$ which is consistent with the condition that $1\ll x<\nu<\eta x$. However, it implies that $\nu^{-1/3}\eta x-\nu^{2/3}=\cO(\nu^0)$. Recalling that $x=2\pi a/\lambda$, this restricts the range of allowed wavelengths. For example, for $a=75\,\mu{\rm m}$, $\nu=1000$, and $0\leq t\leq 10$, we find that $633<\lambda/{\rm nm}<697$. Therefore if we had employed the uniform asymptotic expansion in our calculations, we would have missed 74 of the 83 spectral singularities we found for this case in Section~5 (See Table~\ref{table1}.)

An important outcome of our investigations is that the cylindrical gain medium we consider has effectively no lasing threshold, i.e., we can generate spectral singularities and have the system begin lasing for extremely small values of the gain coefficient. This confirms the validity of our basic motivation for undertaking this project. We expect that the same result holds also for a spherical gain medium. We plan to study the singular gallery modes of samples with a spherical geometry. Finally, we wish to emphasize that singular gallery modes have the unique property of having an effectively infinite quality factor. From the mathematics of spectral singularities, we know that no passive optical material (real optical potential) can possess a zero-width and hence infinite-Q-factor resonance state. Our results show that this is possible if we have an active medium with very small amounts of gain.
\vspace{.3cm}

\noindent \textbf{{Acknowledgments:}} We wish to thank Ali Serpeng\"{u}zel and Aref Mostafazadeh for useful discussions. This work has been supported by the Scientific and Technological Research Council of Turkey (T\"UB\.{I}TAK) in the framework of the project no: 110T611, and by the Turkish Academy of Sciences (T\"UBA).

\end{document}